\documentstyle[preprint,pra,aps]{revtex}

 \input epsf.tex
\begin{document}

\title{Effect of Contact Interfaces on Quantum Conductance of Armchair 
Nanotubes}

\author{{\sc S. Krompiewski}\footnote{) Corresponding author;
Phone: +48-61-8695-126, Fax: +48-61-8684-524, e-mail: \\
stefan@ifmpan.poznan.pl \\ 
\hspace*{0mm} $^{**}$) Presented at the European
Conference {\it Physics of Magnetism}, Pozna\'n, 1-5 July, 2002}
 )  }

\address{ Institute of Molecular Physics, Polish Academy of Sciences, \\
M. Smoluchowskiego 17, 60-179 Pozna\'n, Poland
 \\
}

 \maketitle

\vspace{1cm}
\begin{center}
(Submitted May 9, 2002)$ ^{**}$
\end{center}

\vspace{1cm}

\hspace{9mm} Subject classification: 81.07.De, 85.35.Kt, 73.63.-b, 75.70.Pa

\begin{abstract}
Effect of contact interfaces, between metallic single-wall carbon 
nanotubes (SWCNT) and external electrodes made also
of nanotubes, on the electrical conductance is studied. A tight-binding model 
with both diagonal and off-diagonal disorder, a recursive Green
function technique as well as the Landauer formalism are used. 
The studies are carried out within the coherent transport regime
and are focused on:  (i) evolution from conductance quantization to 
resonant tunneling, (ii) SWCNT's length effects and (iii) magnetoresistance.
It is shown that 
the so-called on-resonance devices, i.e. nanotubes having a conductance peak 
at the Fermi energy, occur with a period of 3 carbon inter-ring spacings.
Additionally, the present approach provides an insight into magnetoresistance dependence
of SWCNTs on conditions at the contact interface.

\end{abstract}


Since carbon nanotubes were discovered in 1991 \cite{Iijima},
a great amount of research has been devoted to their mechanical,
transport and electric properties \cite{book}. Carbon nanotubes are already
used as prototype devices such as field emitters, current
rectifiers or single-electron transistors \cite{phys,set}.
Also from the theoretical view point the carbon nanotubes are a
fascinating subject to study. Their electrical properties depend
critically on the so-called chirality and range from metallic
to semiconducting. The metallic properties are expected provided
that the wrapping vector components
$(n,m)$, defined on the graphene plane, satisfy
the condition $n-m=3 \cdot integer $.  

The system under consideration here is a purely molecular
one, consisting of a central part
made of a single-wall carbon nanotube (SWCNT) sandwiched
between two semi-infinite defect-free lead wires made
also of the SWCNTs. Such molecular systems can now be 
fabricated, with the central segment defined e.g. by creating either
bent or twisted interfaces (see \cite{Orlikowski}, \cite{Rochefort2}
and the references therein). Hereafter 
the central part and the lead wires will be  
referred to as the sample and the electrodes, respectively.
Although in practice molecular systems are coupled with the
macroscopic world via extra metallic electrodes (usually Au), the
latter have negligible effect on the total resistance which
is nearly entirely due to the sample and the interfaces (contacts)
between the sample and the electrodes. So formally the Au electrodes 
may be incorporated into the electron reservoirs and skipped.

The whole system is described by a single-band tight-binding Hamiltonian

\begin{equation} \label{hamiltonian}
\hat H = \sum \limits_{I ,J}
T_{I,J}\vert I><J \vert 
+\sum \limits_{I}
D_I \vert I><I \vert ,
\end{equation}

where $\vert I>$ and $ \vert J> $ stand for orbitals related to
repeat units (principle layers) within both the electrodes and
the sample. So the Hamiltonian has got a block-tridiagonal form,  
with off-diagonal matrices denoted by $T$, and the on-diagonal 
ones $D$.
The theoretical approach applied here -- based on the
Landauer formula and the Green's function method -- proceeds
along the lines of Ref. \cite{Lake}. In particular, the conductance
is expressed in terms of the Green functions and the so-called
broadening functions

\begin{equation} \label{Gamma}
\Gamma^{L,R}=i[\Sigma^{L,R}-(\Sigma^{L,R})^\dagger] ,
\end{equation}

where $\Sigma^L=T_{1,0}g^L(0,0) T_{0,1}$ and
$\Sigma^R=T_{N,N+1}g^R(N+1,N+1) T_{N+1,N}$ with the $Ts$, $g^L$, $g^R$ standing
for the coupling matrices, and the left- and right-electrode Green functions.

In this paper we want to achieve 3 objectives: 
(i) to explain the effect of conditions at sample-electrode
interfaces and (ii)
 of the sample length on the electrical conductance, and (iii) to propose a simplistic 
approach to the spin-polarized transport (giant magnetoresistance, GMR)
through the SWCNT. 
 Letting
the transport remain phase-coherent, it will be tested how the
conductance evolves when the interface conditions change. The conditions
are defined by means of just two parameters, namely the hopping
across the interface, $t_c$ and the on-site potential of the first
carbon ring of the SWCNT sample and the last one.    
For armchair SWCNTs studied here (n=m) the repeat
unit, $a=2.49 \AA$, contains two carbon rings.
The first carbon ring as well as the last one can have
an arbitrary on-site potential, whereas the other
on-site potentials are set to $\epsilon=0$. Similarly, all
the interatomic hopping integrals are $t_{i,j}=-1$
($ \left |t_{i,j} \right |$ is the energy unit),
except for those passing through the interfaces 
 which are also arbitrary, $t_c$.
Figure~1 illustrates the effect of the interface hopping parameter
on the conductance. 

\begin{figure}[h]                           
  \epsfysize=5cm           
  \epsfxsize=4cm           
  \centerline{\epsfbox{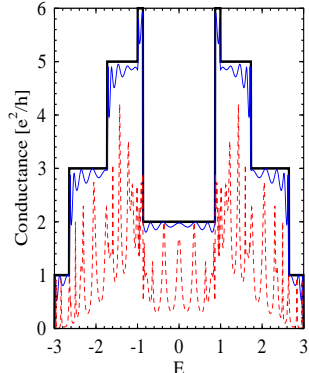}}    
  \caption{
Conductance (per spin) through the armchair SWCNT placed 
between electrodes made also of the SWCNTs (all with $n=m=3$). 
The thick line corresponds to the pristine SWCNT
($t_c=-1$, $\epsilon=0.0$), whereas the solid thin
line and the dashed one correspond to the length
$L=7a$ and the interface
parameters $t_c=-0.9$, $\epsilon=0.0$, and
$t_c=-0.5$, $\epsilon=0.0$, respectively 
 }     
 \label{dos}                     
\end{figure}

It is readily seen that the conductance
evolves gradually from the continuous step-function
quantized at integer multiples of $2 e^2/h$ to the
collection of narrow peaks -- typical of the resonant 
tunneling. The overall picture does not depend on the
tube diameter (determined by $m$) and varies in a 
quasi-periodic way with the sample length (apart from the
trivial case of $t_c=-1$, $\epsilon=0$ corresponding to the
infinite perfect SWCNT). An interesting feature of Fig.1
is the existence of the central peak at energy corresponding
to the Fermi energy E=0. The origin of this has been explained in terms of the so-called 3N+1 rule \cite{Orlikowski}.
If the sample length is equal to $L=(3N+1)a$, with N being an integer number, the allowed discrete values of the 
wave-vector
include k=$2 \pi/3a$ which zeros the electron dispersion relation.
When the amount of disorder increases, the wave-vector is no longer
a good quantum number, so there is no reason for the 3N+1 rule
to be still valid. Fig.~2, plotted for E=0 ant $t_c=-0.5$, shows
it very clearly since the "on-resonance" peaks 
happen to depend strongly on the interface on-site parameters, and in fact 
they may also be described by the 3N rule 
(diamonds) or the 3N-1 rule (triangles).

\begin{figure}[h]                           
  \epsfysize=5cm           
  \epsfxsize=4cm           
  \centerline{\epsfbox{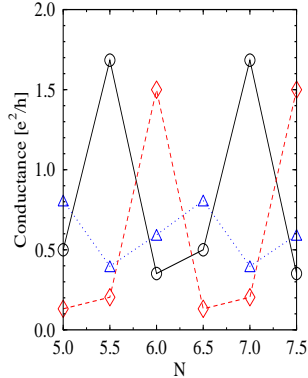}}    
  \caption{
 Length-dependent conductance (per spin) of the SWCNT for E=0 
and various $\epsilon$ parameters. Depending on these parameters,
the so-called on-resonance devices
occur for the length $(3N+1)a$  ($\epsilon=0$, circles), $3Na$ ($\epsilon=1$, diamonds)
and $(3N-1)a$ ($\epsilon=0.5$, triangles), respectively with a period of
$1.5a$
 }     
 \label{dos}                     
\end{figure}

Our calculation method makes it possible to sample the lengths with the resolution of the inter-ring length $a/2$ (half the
unit cell). Owing to this we have found that the period of 
conduction oscillations is $ 3 a/2 $ rather than $3a$. Similarly
as found for finite isolated (not coupled to any electrodes)  SWCNTs \cite{Rochefort1} in the contest
of their energy spectra. 
 It is noteworthy that in the case
of the on-resonance devices the conductance peak height does
not vanish with decreasing $ t_c $ (only its width does)
and tends to the constant value.

    Another point addressed in this paper is the giant 
magnetoresiatance of a single-molecule system. It has been
shown experimentally that carbon nanotubes electrically
contacted by ferromagnetic
cobalt show the GMR effect up to 9\%
\cite{Tsukagoshi}. The effect could be possibly even higher if there
were single-domain electrodes and interfaces of better quality.
To gain a closer insight into this we mimic spin-polarized electron
injection to the SWCNT by applying fictitious magnetic electrodes
which guarantee optimal matching of the interfaces,
making it possible to single out the magnetic contribution. 
The fictitious electrodes have been
constructed from the SWCNT electrodes by simply shifting, 
in a rigid way, the up- and down-spin
 bands with respect to each other. The GMR is defined as

\begin{equation} \label{GMR}
GMR=(\cal {G}_{\uparrow,\uparrow} - \cal {G}_{\uparrow,\downarrow}) /  \cal {G}_{\uparrow,\uparrow} ,
\end{equation}

where $\cal G_{\sigma,\sigma'}$ denotes 
the conductance for the parallel and antiparallel configurations.
N.B. the GMR problem in the SWCNTs
has already been studied before by means of a similar approach
in Ref. \cite{Mehrez}, where the electrodes have been 
treated as featureless leads which are taken into account indirectly 
by parametrizing spin-dependent self-energies in terms of corresponding
line-width functions $ \Gamma_{\uparrow,\downarrow} $ with
a fixed ratio $ \Gamma_\uparrow / \Gamma_\downarrow=2 $. 

Fig.~3 represents the GMR dependence on the spin polarization of injected
electrons for some typical interface parameters $\epsilon$ and $t_c$, 
 and the sample lengths $N=6a$ and $7a$.
It is interesting to note
that for the on-resonance device the {\it inverse} GMR
may occur (thick solid curve), however it gives way to the positive GMR
if some amount of diagonal disorder is present (thin solid line). As regards the
off-resonance devices, their GMR is always positive (dashed curve). The reason
for this peculiar behavior is that the density of states at the Fermi energy depends
strongly both on the length and on conditions at the interfaces. 

\begin{figure}[h]                           
  \epsfysize=5cm           
  \epsfxsize=4cm           
  \centerline{\epsfbox{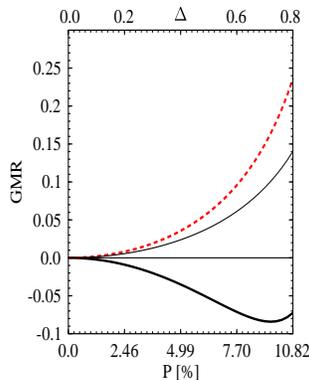}}    
  \caption{
 Giant magnetoresistance of the SWCNT coupled to
fictitious electrodes, against their spin-polarization, P (lower x-axis) and the
rigid band splitting, $\Delta$  (upper x-axis). The parameters are as follows: 
 $L=6a$, $t_c=-0.5$, $\epsilon=0$ (dashed line);
$L=7a$, $t_c=-0.5$, $\epsilon=0$ (thick line);
$L=7a$, $t_c=-0.5$, $\epsilon=0.2$ (thin line).
 Note that the plot is restricted to the range where P scales roughly linearly with $\Delta$,
in order to avoid electrode-specific features
 }     
 \label{dos}                     
\end{figure}

In conclusion, quality of the sample-electrode interfaces
is decisive for transport properties of single-wall carbon
nanotubes. When the amount of disorder at the
interfaces increases, the SWCNT conductance evolves from
the continuous step function (discretized at integer multiples
of $\frac{e^2}{h}$) to the set of peaks typical of the
resonant tunneling. The conductance of armchair SWCNT
oscillates with the period of $3a/2$ ($a$ is the repeat unit cell). 
The on-resonance device is characterized by the existence
of the conductance peak just at the Fermi energy. The on-resonance
device length has been shown here to depend on the interface parameters,
and not to satisfy the 3N+1-rule in general.
The GMR effect in the SWCNT may be negative in the case
of the on-resonance devices with no on-site interface disorder. 

{\it Acknowledgments } This work was partially supported by the grant
No.~PBZ/KBN-013/T08/23. 
I thank the Pozna\'n Supercomputing and Networking Center for the
computing time.

\newpage
\listoffigures

\end{document}